\def\version{31-7-2000}
\def\address#1{\date{{\sl#1}\\\ \\\version}\gdef\date##1{}}%
\documentclass[twoside,11pt]{article}
\usepackage{cite,amsfonts,a4wide}

\def\openone{\leavevmode\hbox{\small1\kern-3.8pt\normalsize1}}%
\def\dim{\mathop{\rm dim}\nolimits}
\def\tr{\mathop{\rm Tr}\nolimits}
\def\acknowledgements{\section*{Acknowledgements}}%

\makeatletter
\@addtoreset{equation}{section}
\makeatother

\def\g{{\mathfrak g}}
\def\ie{{\sl i.e.\/}}
\def\eg{{\sl e.g.\/}}

\def\etc{{\sl etc.\/}}
\def\cf{{\sl cf.\/}}
\def\Cf{{\sl Cf.\/}}

\newcommand{\bb}{\begin {minipage} {3cm}\begin{center}}
\newcommand{\ee}{\end{center}\end{minipage}}
\newcommand{\bc}{\begin {minipage} {2.5cm}\begin{center}}
\newcommand{\bd}{\end{center}\end{minipage}}

\newcommand {\R} {{\mathbb R}}

\newcommand {\q}{\begin{quote} \small}

\newcommand {\be}{\begin{equation}}
\newcommand {\e}{\end{equation}}
\newcommand {\bea}{\begin{eqnarray}}
\newcommand {\ea}{\end {eqnarray}}
\newcommand {\cl}[1]{{\cal #1}}
\newcommand {\bo}[1]{{\bf #1}}
\newcommand {\frack}[2]{\mbox{${\textstyle{\frac{#1}{#2}}}$}}

\begin{document}
\title{
 \flushright{{\sl{DAMTP-2000-64}}}  \\ ${}$ \\
Symplectic and orthogonal Lie algebra technology for bosonic \\ 
  and fermionic oscillator models of integrable systems $\qquad$} 
\author{
  A.~J.~Macfarlane\thanks{e-mail: A.J.Macfarlane@damtp.cam.ac.uk},
  Hendryk Pfeiffer\thanks{e-mail: H.Pfeiffer@damtp.cam.ac.uk}\ \ and 
  F.~Wagner\thanks{e-mail: F.Wagner@damtp.cam.ac.uk}}
\address{Centre for Mathematical Sciences,\\
  Department of Applied Mathematics and Theoretical Physics,\\
  Wilberforce Road, Cambridge CB3 0WA, UK} 
\date{\version}
\maketitle

\begin{abstract}
To provide tools, especially $L$-operators,
for use in studies of rational Yang-Baxter algebras and
quantum integrable models when the Lie algebras $so(N)$ ($b_n$, $d_n$)
or $sp(2n)$ ($c_n$) are the invariance algebras of their $R$ matrices,
this paper develops a presentation of these Lie algebras convenient
for the context, and derives many properties of the matrices of their
defining representations and of the ad-invariant tensors that enter
their multiplication laws. Metaplectic-type representations of $sp(2n)$ and 
$so(N)$ on bosonic and on fermionic Fock spaces respectively
are constructed. Concise general expressions (see (\ref{5.2}) and (\ref{5.11})
below) 
for their $L$-operators are 
obtained, and used to derive simple formulas for the 
$T$ operators of the rational $RTT$ algebra of the associated integral systems,
thereby enabling  their efficient treatment by
means of the algebraic Bethe ansatz.
\end{abstract}
%
%
\section{Introduction}
%

Let the compact real simple Lie algebra $\g$ with generators $X_i$ and
totally antisymmetric structure constants be defined by
\begin{equation} 
\label{1.1} 
[X_i \, , \, X_j] =ic_{ijk} X_k \quad. 
\end{equation}
Let $\cl{V}$ denote the defining representation of $\g$ with matrices
$x_i$ given by $X_i \mapsto \gamma x_i$, where $\gamma=1$
for all series of simple Lie algebras  except
$a_n=su(n+1)$, for which $\gamma=\frac{1}{2}$.  We chose the basis so
that
\begin{equation} 
\label{1.2} 
  {x_i}^{\dagger}=x_i,\quad\tr x_i=0,\quad\tr x_ix_j=2\delta_{ij}\quad. 
\end{equation}
We define also the $L$-matrix by
\begin{equation} 
  \label{1.3} 
  L=x_i \otimes X_i \equiv x_i X_i
\end{equation} 
acting on $\cl{V}\otimes\cl{H}$ where $\cl{V}$ is the defining
representation of $\g$ and $\cl{H}$ any other representation.  This
matrix is an object of major importance for work on integrable systems
defined with the aid of the rational Yang Baxter algebra on 
$\cl{V}\otimes \cl{V}\otimes \cl{H}$ 
\begin{equation} 
  \label{1.4} 
  R_{12}(u-v) T_1(u) T_2(v)=T_2(v) T_1(u) R_{12}(u-v) \quad , 
\end{equation}
for which $\g$ is the Lie algebra under which the corresponding rational $R$
matrix is invariant.  Here the subscripts $1$ and $2$ refer to the
auxiliary vector spaces $\cl{V}$, and the $T_i(u)$ act on
$\cl{V}\otimes\cl{H}$ where $\cl{H}$ is the quantum space. The
$T_i(u)$ are defined in terms of the $X_i$, and may be used to
provide the observables of the integrable
system under study. In fact, one uses $L$ in the construction of
one-site solutions of $T(u)$ of (\ref{1.4}), and then follows the
procedure laid down by Sklyanin~\cite{skly}, to build $N$-site
integrable models of spin chain type based on the Lie algebra $\g$.

In the study of integrable quantum mechanical models, it is widely
accepted that much deeper insight into their structures -- why things
work as they do -- is attained by generalising from some simple basic
model, in which usually $\g=su(2)$, to its counterpart for general
$\g$. Generalisation from $a_1=su(2)$ to $a_{n-1}=su(n)$ is normally the
easiest to achieve, and in part this is due to the fact that much more
detailed and systematic knowledge is available in this case than is 
conveniently at  hand for $c_n=sp(2n), b_n=so(2n+1)$ and $d_n=so(2n)$.

The present paper originated in the study~\cite{MW} of integrable
systems in which the quantum space is the representation space of the
metaplectic representation~\cite{Itz,SW} of $c_n$ for which there is
an elegant realisation in terms of $n$ pairs of (bosonic) oscillator
variables. The system in question derives its interest in large part from the 
fact that it is one for which the determination of its physical eigenstates can
be carried out, for $\g$ `larger' than $a_1$, by an unnested
application of the algebraic Bethe ansatz
procedure~\cite{FAD,GraS}. (\Cf\ the treatment of~\cite{FAD,GraS} with
that of~\cite{Resh1} for $g=su(3)$, or of \cite{Resh3} and \cite{deV}, 
which give rise to
nested Bethe equations for the determination of the physical
eigenstates of the models studied.)

Now the metaplectic realisation of $c_n$ in terms of bosonic
oscillators has a close analogue at a fairly deep structural
level in which the spinor representations of $d_n=so(2n)$ are realised
in terms of $n$ pairs of Dirac fermionic operators. Just as the
Hilbert space of the metaplectic representation of $c_n$ consists of
two irreducible subspaces of even and odd states, so also does the
fermionic construction give rise to the two spinor representations of
$d_n$ of opposite chirality, carried by the subspaces of even and odd
total fermion number. A deep treatment of this situation is contained
in \cite{segal}.
In general what can be done for one of the two types of  
theories admits a parallel treatment in the other. It is perhaps worth 
remarking that we here use the word metaplectic 
in the way that is usual in theoretical physics, 
primarily in relation to the $c_n$ context. It is used more
generally in the mathematical literature, as can be traced from \cite{nek}.
 
Having discovered the relevance of the fermionic realisation of $d_n$
in the context of integrable systems, it was natural to consider the 
extension of our
results to $b_n$.  This is easily achieved by adjoining a Majorana
fermion to the variables of the realisation of $d_n$.  However the
need to take account of this renders awkward the Fock space treatment
of the spinor representation of $b_n$ that has been realised, making
related models less favourable candidates for integrable
system study. In this context, we observe that in the $c_n$ and $d_n$
examples, there is an obvious relationship between the 
models we build and the Lie algebra decompositions associated
with the hermitian symmetric spaces~\cite{helg} $Sp(2n,{\R})/U(n)$ and
$SO(2n)/U(n)$. There is no such relationship in the case of our $b_n =
so(2n+1)$ examples.
 
To account for our interest in the two `metaplectic' type theories, we
remark that they are very favourably placed amongst the larger family
of algebraic systems (see~\cite{gun} for a review and~\cite{vdJ} for
an interesting example) that have nice descriptions in terms of
oscillators. At the simplest level of construction, the ansatz
\begin{equation} 
  \label{1.5}
  X_i= \gamma A^\dagger x_i A \quad, 
\end{equation}
produces an operator realisation of $\g$ whenever the column vector
$A$ has either bosonic or fermionic creation operators as its set of
$\dim\, \cl{V}$ components. Proof in each case is
elementary. Roughly speaking, the two types of model that we are focusing our
attention upon use only {\it half} as many pairs as bosonic or fermionic
variables as
eq.~(\ref{1.5}) does. To see this it is simplest to look ahead to
eq.~(\ref{4.2}) for $c_n$ and~(\ref{4.12}) for $d_n$ where a single
vector corresponding to $A$ contains all creation and annihilation
operators of the system. The advantages are  not only that we have smaller Fock
spaces to deal with and thus a gain in tractability, but also that 
specially simple properties of
the operators $L$ and $T(u)$ come into play, which allow considerable
simplifications. In particular, we obtain strikingly concise general formulas,
valid for all $n$,
for the $L$-operators, see (\ref{5.2}) for $c_n$ and (\ref{5.11})
for $d_n$ below. These expressions make it very easy to show that
$L$ then obeys a quadratic equation. It is this property which
makes the solution $T(u)$, displayed below as   eq.~(\ref{t_def}) 
for a one-site solution of
eq.~(\ref{1.4}), take on its particularly simple form.
Another construction which, when it is compared to (\ref{1.5}),
in some sense shares the economy of the models we are talking about now, uses a
vector of hermitian {\it Majorana} fermions. For one example of this,
see~\cite{ajmmf}; for a more recent one, see \cite{dAMf}. Indeed,
if the fermionic
models of Sec. 4.2 are formulated in terms of Majorana fermions, their
relationship to models like those of \cite{ajmmf} and \cite{dAMf} 
can be seen to be quite close.

Our continuing studies of these problems have called for the
systematic assembly of information about $c_n=sp(2n)$ and $so(N)$
($b_n$ and $d_n$). One reason why the work described here is claimed
to be novel concerns the description we give of the matrices $x_i$ of
the defining representations $\cl{V}$ of these algebras.  We have
found it best to use a set labelled by a single index, rather than the
more common alternative in which one uses, for example for $so(N)$, a
basis in $\cl{V}$ given by $x_{ij}= -x_{ji}= (E_{ij}-E_{ji})/2$, where
$E_{ij}$ has as its only non-zero element a one where row $i$ meets
column $j$.  Our choice arises mainly because the physical description
of the models we are interested in favours a Cartan-Weyl basis for $\g$,
and the two-index labelled basis for $\g$ does not relate very well 
to this. One consequence of our choice of 
basis is that
it brings into the formalism ad-invariant analogues of the $f$ and $d$
tensors of $a_{n-1}=su(n)$.  One of the benefits is seen in the ease
with which one accumulates all the necessary data about the Lie
algebras, representations of them including the ones important to our
programme, and the associated $L$ matrices.

The paper is organised as follows. Sec. 2.1
contains all the necessary information about the description of the
Lie algebras $c_n$, $b_n$ and $d_n$ in Cartan-Weyl form. It is
presented so that as much of the data as possible is encapsulated in
the $L$ matrix of the Lie algebra. It turns out that we can 
present the matrix $L$ in an essentially standard form whose main features are
determined by the way the positive roots are related to the simple
ones: Sec. 2.1  aims at explaining this carefully. One says
`essentially' because the pattern observed for $c_n$ and $b_n$, and
available (and very simply) also for $a_n$, and even $g_2$, is not
visible immediately in the corresponding $L$ for $d_n$.  The
features of $d_n$ which obscure the simple pattern seen in other cases
are associated with the fishtail extremity of its Dynkin
diagram. However, by viewing the two equivalent fishtail-roots of the
diagram suitably, the $d_n$ pattern can be made to conform
satisfactorily to that enjoyed by the other Lie algebras studied here.

Sec. 3 derives a compendium of results involving
the matrices $x_i$ of $\g$ for $c_n$ and, in unified form, for
$so(N)$, certain associated matrices $y_\alpha$, and the various
ad-invariant tensors that enter naturally into the product laws of
these matrices. The matrices $y_\alpha$ enter necessarily alongside
the $x_i$ to complete the basis of the space of all traceless
$2n\times 2n$ hermitian matrices for $c_n$ and $N\times N$ for $so(N)$. The
treatment perhaps goes further than the needs of this paper make apparent.
Even so it is certainly not comprehensive, although it should
indicate viable methods of approaching further identities.

Sec. 4  describes the bosonic oscillator
description of the metaplectic representation of $c_n$, and 
compares and contrasts this with the fermionic
analogues for $d_n$ and $b_n$, giving also some discussion of all the
corresponding Fock spaces.

Sec. 5 discusses various matters associated with the fact that the 
$L$-operators dealt with in this paper obey quadratic equations.
It begins by displaying concise and tractable formulas for the 
$L$ operators of each of the oscillator systems of Sec. 4, and thence 
derives the important quadratic equations
satisfied by these $L$-operators, mentioning also some $a_n$
examples based on (\ref{1.5}) for the purposes of comparison. Also it
is shown how the equations for $L=x_i X_i$ can be used to provide
product laws for the matrices/operators $X_i$ involved in $L$. There
are new results here. They are simple because $L$ obeys a quadratic
equation. But one could in principle seek similar results -- for
$X_{(i} X_j X_{k)}$ -- whenever $L$ obeys a cubic equation, and
beyond. Also the practical relevance to the solution of the corresponding 
$RTT$ equations (\ref{1.4}) of the fact that $L$-operators obey 
quadratic equations is indicated briefly. Further, the correlation between 
that property of such an $L$-operator acting in a space 
$\cl{V}\otimes\cl{H}$ ({\it cf.} (1.3)$\;$), and the fact  
that this space reduces, under the action of the relevant invariance algebra 
$\g$, to a direct sum of exactly two subspaces, in each of which the quadratic 
Casimir $C^{(2)}_{\cl{V}\otimes\cl{H}}$
of $\g$ takes on a fixed value, is noted, and a survey is given of the
situations in which such a reduction takes place.

%
\section{Cartan-Weyl form of the defining representations}
\label{sec_cw}
%

We present the Lie algebra $\g$ of rank $n$ and $\dim\,\g$ in 
Cartan-Weyl form with generators $\bo{H}
=(H_1,\ldots,H_n)$ of its Cartan subalgebra, positive roots
$\bo{r}_\alpha$, and raising and lowering operators $E_{\pm \alpha}$,
$\alpha \in \{ 1,2,\ldots,\frack{1}{2} (\dim\,\g-n)\}$.  Then we
have
\be \label{2.1}
  [\bo{H},E_{\pm\alpha}] \, = \,  
\pm\bo{r}_\alpha E_{\pm \alpha} \quad , \quad 
  [E_{\alpha},E_{-\alpha}]= \bo{r}_\alpha\cdot\bo{H} \quad, 
\e  
together with well-known non-trivial expressions for $[E_{+\alpha} \,
, \, E_{\pm \beta} ]$ whenever $\bf{r}_{\alpha} \pm \bf{r}_{\beta}$ is
a non-zero root of $\g$. The $X_i$ of (\ref{1.1}) are related to the
Cartan-Weyl generators according to
\be \label{2.2}
   \Bigl\{  X_i\colon i\in\{1,\ldots,\dim\g\}\Bigr\} 
    = \Bigl\{ \begin{array}{l}  H_r\colon r \in \{1,\ldots,n\} \\
          U_\alpha \, , \, V_\alpha\colon \alpha \in \{1,\ldots,
       \frack{1}{2}(\dim\,\g-n)\} \end{array} \Bigr. \quad  ,
\end{equation}
where $\sqrt{2} E_{\pm \alpha}=U_\alpha\pm i V_\alpha $. For the
defining representation $\cl{V}$ of $\g$, we have $X_i \mapsto \gamma
x_i$ with $\gamma=1$, ( except for $a_n$ when
$\gamma=1/2$) so that we also employ the notation
\begin{equation} 
\label{2.3} 
  \bo{H}\mapsto\bo{h}, \quad E_{\pm \alpha}\mapsto e_{\pm \alpha},\quad
  \sqrt{2 }e_{\pm \alpha}= u_{\alpha} \pm i v_{\alpha}\quad. 
\end{equation}
Further, for compact real $\g$, we have ${x_i}^{\dagger}=
x_i$ for all $i$, and hence $\bo{h}^{\dagger}=\bo{h} \, , \, 
{e_{+\alpha}}^\dagger=e_{-\alpha} $, but no such result will hold for a 
non-compact representation of $\g$.
We now turn case by case to the definition of $L$-matrices 
for the Lie algebras $\g$ of immediate interest, intending to comment below
on the nearly standard form of the results displayed.

General references for background on Lie algebras and their
representations are~\cite{hum,FS,corn}. Another source of valuable
information is~\cite{slan}.

\subsection{The Lie algebras $c_n$}
\label{sec_cn}

We begin with the
case of $c_2 = sp(4)$, but everything generalises naturally for
$c_n = sp(2n)$. For $c_2$, the simple roots are $\bo{r}_1=(1,-1)$
and $\bo{r}_2=(0,2) \;$ (see~\cite{hum}: p64). The remaining positive
roots are $\bo{r}_3=\bo{r}_1+\bo{r}_2 \equiv \bo{r}_{12}=(1,1)$ 
and $\bo{r}_4=\bo{r}_1+\bo{r}_3=2\bo{r}_1+\bo{r}_2 \equiv \bo{r}_{112}
=(2,0)$. Hence, in addition to (\ref{2.1}), we have
\begin{equation} 
\label{2.5} 
  [E_1,E_2]=\sqrt{2}E_3\quad,\quad 
  [E_1,E_3]=\sqrt{2}E_4\quad,
\end{equation}
together with results obtained from these in a well-known manner. 
In this basis, $L$ from eq.~(\ref{1.3}) is
of the form
\begin{equation} 
\label{2.6} 
  L = h_1 H_1 + h_2 H_2 
    + \sum_{\alpha=1}^4( e_{+\alpha} E_{-\alpha} + e_{-\alpha}
    E_{+\alpha} )\quad. 
\end{equation}
If we write the left tensor factor
as matrices and the right one as operators, so that $L$ reads
\begin{equation} 
\label{2.7} 
\left(
\begin{array}{rrrr}
H_1 & E_{-1} & E_{-3} & \sqrt{2} E_{-4} \\
E_1 & H_2    &  \sqrt{2} E_{-2} & E_{-3} \\
E_{3} & \sqrt{2} E_2 & -H_2 & -E_{-1} \\
\sqrt{2} E_4 & E_{3} & -E_1 & -H_1 
\end{array} 
\right)\quad,
\end{equation}
then one can read the explicit forms of the matrices $h_1$, $h_2$ and
$e_{\pm \alpha}$ off (\ref{2.7}) and verify that they do indeed
provide a representation of the Lie algebra relations of $c_2$,
obtained from (\ref{2.1}) and (\ref{2.5}).  Furthermore, and in part
because of the $\sqrt{2}$ factors (which multiply the raising
generators which belong to the \emph{long} roots of $c_2 \;$), the
$x_i$ clearly obey (\ref{1.2}), and also
\begin{equation} 
\label{2.8}
  {x_i}^T= -J x_i J^{-1} \quad ,
\end{equation}
where $J$ is the standard symplectic form, which in our representation
(and in~\cite{GR}) is given by
\begin{equation} 
\label{2.9} 
\left(
\begin{array}{rrrr}
0 & 0 & 0 & 1 \\
0 & 0 & 1 & 0 \\
0 & -1 & 0 & 0 \\
-1 & 0 & 0 & 0 
\end{array} \right) \quad .
\end{equation}
We turn next to $c_3$. Here the symplectic form $J$ has, as its only 
non-zero entries,\\ $(1,1,1,-1,-1,-1)$ down the main
antidiagonal. The 
simple roots of $c_3$, as given $\in {\R}^3$ (\cite{hum}: p64) are 
\begin{equation} \label{2.11}
\bo{r}_1=(1,-1,0) \; , \; \bo{r}_2=(0,1,-1) \; , \; \bo{r}_3=(0,0,2) \quad ,
\end{equation}and other positive roots are given by
\begin{eqnarray}  \label{2.12}
\bo{r}_4=\bo{r}_1+\bo{r}_2 \equiv \bo{r}_{12} \; , \; & \bo{r}_5=\bo{r}_2+
\bo{r}_3 \equiv \bo{r}_{23} & \; , \; 
\bo{r}_6= \bo{r}_1+\bo{r}_2+\bo{r}_3 \equiv \bo{r}_{123}  \quad ,
\nonumber \\
\bo{r}_7=2 \bo{r}_2+\bo{r}_3 \equiv \bo{r}_{223} \, , & \bo{r}_8= 
\bo{r}_1+2\bo{r}_2+\bo{r}_3 \equiv \bo{r}_{1223} & , \, \bo{r}_9=2\bo{r}_1+
2\bo{r}_2+\bo{r}_3 \equiv \bo{r}_{11223}
\, .
\end{eqnarray} 
Given these the requisite form of $L$ for $c_3$ can directly be inferred to be
\begin{equation} 
\label{2.10} 
\left(
\begin{array}{rrrrrr}
H_1 &&&&& \\
E_1 & H_2 &&&& \\
E_{12} & E_2 & H_3 &&& \\
E_{123} & E_{23}  & \sqrt{2} E_3 &-H_3 && \\
E_{1223} & \sqrt{2} E_{223} & E_{23}  & -E_2 & -H_2 & \\
\sqrt{2} E_{11223} & E_{1223} & E_{123} & -E_{12} & -E_1  & -H_1
\end{array} \right) \quad. 
\end{equation}
For clarity, we left out the entries involving the $E_{-\alpha}$ that
should occupy the places above the main diagonal of (\ref{2.10}).
We have moreover employed a notation that at first glance might seem unduly
clumsy in order to make evident some features of our construction of $L$ that
we return to below. We can however immediately read off (\ref{2.10}) 
explicit expressions for the $6 \times 6$ matrices of the defining 
representation $\cl{V}$ of $c_3$, namely $h_1$, $h_2$, $h_3$
and $e_{\pm \alpha}$ for $\alpha \in
\{1,\ldots,9\}$, or, assigned in accordance with (\ref{2.11}) and (\ref{2.12}),
for 
\be \label{2.al}
\alpha \in \{ 1, 2, 3, 12, 23, 123, 223, 1223, 11223 \} \quad . \e
Then the Lie algebra
of $c_3$ can be calculated  from the matrices of  $\cl{V}$, and seen to be of 
Cartan-Weyl form with roots given correctly by (\ref{2.11}) and (\ref{2.12}).
In other words $L$ as given by (\ref{2.10}) is the Cartan-Weyl form of 
$L=x_i \, X_i$, wherein the matrices $x_i$ of $\cl{V}$ obey the same algebra
(\ref{1.1}) as the abstract generators $X_i$.

We wish next to state in full all the steps of the recipe that enables the 
above construction to be applied to $c_n$ of arbitrary rank $n$, and beyond.
Consider first the placement of the $E_{\pm \alpha}$ in (\ref{2.10})
in relation to the root system of $c_3$. It can be
seen to follow a
systematic pattern, that will allow the exact form of $L$ to be written down
directly for {\it arbitrary} $c_n$. The form of $L$ displayed above for
$c_2$ in (\ref{2.7}) of course conforms to the same pattern although
we did not stop there to point this out.  The patterns are 
easiest to see if one
writes (\ref{2.7}) schematically, displaying, at the site occupied by a
given raising generator, only the corresponding root label
\begin{equation} \label{2.12A} \left(
\begin{array} {cccc}
x &&& \nonumber \\
1 & x && \nonumber \\
12 & 2 & x & \nonumber \\
112 & 12 & 1 & x
\end{array} \right) \quad , \end{equation}
and similarly (\cf\ (\ref{2.10})) for $c_3$
\begin{equation} \label{2.12B} \left(
\begin{array} {cccccc}
x &&&&& \nonumber \\
1 & x &&&& \nonumber \\
12 & 2 & x &&& \nonumber \\
123 & 23 & 3 & x && \nonumber \\
1223 & 223 & 23 & 2 & x &  \nonumber \\
11223 & 1223 & 123 & 12 & 1 & x  
\end{array} \right) \quad . \end{equation}

Since inspection of these displays surely makes plain their salient features,
we now state our recipe in full. First a natural choice of Cartan subalgebra 
generators is made
and they are placed down the main diagonal. Second the
raising generators corresponding to the simple roots are placed down
the first sub-diagonal. The symmetry present after performing the first step 
dictates how to perform the second one.
Third, the positive root sums determine the rest
of the assignments according to the uniform pattern implicit in (\ref{2.12A})
and (\ref{2.12B}). Explicitly then consider {\sl e.g.} the entry to 
(\ref{2.10}) that sits in the place $L_{51}$, and look 
at the portion of the first sub-diagonal subtended by $L_{51}$. We find there,
occupying the places $L_{54},L_{43},L_{32},L_{21}$, the entries
$-E_2,\, \sqrt{2}E_3,\, E_2,\, E_1$. Disregarding the minus sign and the 
root-two, these entries correspond to simple roots whose sum is
$\bo{r}_{1223}$, and we therefore associate $E_{1223}$ with the place $L_{51}$,
so that the stage summarised in (\ref{2.12B}) is reached.
Fourth, while we have disregarded the signs of elements $L_{ij}$ for
the \emph{identification} of what raising operators to assign to 
places in $L$ below the
first sub-diagonal, these signs have to be supplied.
This is simple to do because the pattern of signs in $L$
possesses an evident and uniform pattern, trivial here and similarly
for the case of $b_n$ treated below. These patterns stem 
directly from our use of Racah metric forms, like (\ref{2.9}) here and 
(\ref{2.22}) below for $b_n$,
in determining transposition properties, (\ref{2.8}) here and (\ref{2.21}) 
below for $b_n$ and $d_n$,
of the matrices $x_i$ of $\cl{V}$. 
Fifth, factors $\sqrt{2}$ are inserted as needed in order to normalise the 
matrices $x_i$ so that $\tr x_i\, x_j=2\delta_{ij}$. It is easy to check the 
trace and transposition properties of the matrices of $\cl{V}$ read off 
displays like (\ref{2.20}) by considering the matrix $A=a_i x_i$ for $a_i \in
\R$ for which ${\rm Tr}\, A^2=a_i a_i$ and $A^T=-JAJ^{-1}$.

A state of affairs analogous to what has just been described is observed, below
straightforwardly for $b_n$ and  
(in a considerably more subtle form) for $d_n$, and found
elsewhere~\cite{ajmg2} also for $g_2$.  

Our purposes here do not require us to
address questions regarding the $a_n$ family, but it is easy
to present $L$ for this family in a way that embodies very simply the
features in focus just now. See the book~\cite{FH} which uses matrices
like $L$ in a fashion that is quite similar to what is done
here. Similarly see~\cite{nek}. We turn next to the case of $b_n$.

\subsection{The Lie algebras $b_n$}

In this case, $b_n = so(2n+1)$, it is convenient to present results
for $n=2$. Despite the fact that $b_2 \cong c_2$, the results for
$n=2$ do make completely evident the full generalisation for all
$n$. The simple and positive roots are in this case chosen to be
$\bo{r}_1=(1,-1)$, $\bo{r}_2=(0,1) \;$ \cite{hum}, $\; \bo{r}_3=
\bo{r}_1+\bo{r}_2 \equiv \bo{r}_{12}=(1,0)$ and $\bo{r}_4=\bo{r}_1 +
2\bo{r}_2\equiv \bo{r}_{122}=(1,1)$, and we have
\begin{equation} \label{2.20} \left(
\begin{array}{rrrrr} 
H_1 & E_{-1} & E_{-3} & E_{-4} & 0 \\
E_{1} & H_2 & E_{-2} & 0 & -E_{-4} \\
E_{3} & E_{2} & 0 & -E_{-2} & -E_{-3} \\
E_{4} & 0  & -E_{2} & -H_2 & -E_{-1} \\
0 & -E_{4} & -E_{3} & -E_{1} & -H_1 
\end{array} \right) \quad .
\end{equation}
Again all of $h_1$, $h_2$ and the $e_{\pm \alpha}$ can be read off
(\ref{2.20}) and shown to obey the correct Lie algebra relations. Here
the corresponding $5 \times 5$ matrices $x_i$ satisfy (\ref{1.2}) and
\begin{equation} \label{2.21}
{x_i}^T= -Mx_iM^{-1} \quad ,
\end{equation} where $M$ is the Racah~\cite{GR} form for $b_2$
\begin{equation} \label{2.22} \left(
\begin{array}{ccccc} 
0 & 0 & 0 & 0 & 1 \\
0 & 0 & 0 & 1 & 0 \\
0 & 0 & 1 & 0 & 0 \\
0 & 1 & 0 & 0 & 0 \\
1 & 0 & 0 & 0 & 0 
\end{array} \right) \quad .
\end{equation}
The evident generalisation to $b_n$ ( for which the \emph{short} roots
are $E_3$, $E_{23}$ and $E_{123} \;$) gives for $n=3$
\begin{equation} \label{2.23} \left(
\begin{array}{ccccccc} 
H_1 &&&&&& \\
E_1 & H_2 &&&&& \\
E_{12} & E_2 & H_3 &&&& \\
E_{123} & E_{23} & E_3 & 0 &&& \\
E_{1233} & E_{233} & 0 & - E_3 & -H_3 && \\
E_{12233} & 0  & -E_{233} & -E_{23} & - E_2 & \phantom{x}  -H_2 & \\
0  & -E_{12233} & -E_{1233} & -E_{123} & -E_{12} & 
\phantom{x} - E_1 & \phantom{xx} -H_1 
\end{array} \right) \quad .
\end{equation}
Here the simple roots are~\cite{hum}
\begin{equation} \label{2.24} \bo{r}_1=(1,-1,0) \; , \; 
\bo{r}_2=(0,1,-1) \; , \;
\bo{r}_3=(0,0,1) \quad , \end{equation}
and the remaining positive roots can be inferred from (\ref{2.23}) to be 
given by
\begin{eqnarray} \label{2.25}
\bo{r}_{12}=\bo{r}_{1}+\bo{r}_{2} &=& (1,0,-1)  \quad , \nonumber \\
\bo{r}_{23}=\bo{r}_{2}+\bo{r}_{3} &=& (0,1,0)  \quad , \nonumber \\
\bo{r}_{123}=\bo{r}_{12}+\bo{r}_{3}=\bo{r}_{1}+\bo{r}_{2}+\bo{r}_3 &=& 
(1,0,0)  \quad , \nonumber \\
\bo{r}_{233}=\bo{r}_{23}+\bo{r}_{3}=\bo{r}_{2}+2\bo{r}_3 &=& 
(0,1,1)  \quad , \nonumber \\
\bo{r}_{1233}=\bo{r}_{123}+\bo{r}_{3}=\bo{r}_1+\bo{r}_{2}+2\bo{r}_3 &=& 
(1,0,1)  \quad , \nonumber \\
\bo{r}_{12233}=\bo{r}_{1233}+\bo{r}_{2}=\bo{r}_1+2\bo{r}_{2}+2\bo{r}_3 &=& 
(1,1,0) \quad .
\end{eqnarray} 
One can also read from (\ref{2.23}), whose upper triangular part
involving the lowering generators has been suppressed, explicit
matrices $h_1, h_2, h_3$ and $e_{\pm \alpha}$ where $\alpha$ runs
through the set of values
\begin{equation} 1, 2, 3, 12, 23, 123, 233, 1223, 12233. \end{equation}
Then one can check that they satisfy the correct Lie algebra
relations, with the same expressions for the roots as given in
(\ref{2.24}) and (\ref{2.25}).  We could also have presented the $b_2$
results in such a fashion using notation for $b_2$ like $E_3=E_{12}$
since for $b_2$ we have $\bo{r}_{3}=\bo{r}_{1}+\bo{r}_{2} \equiv
\bo{r}_{12}$, etc.  The display (\ref{2.23}) should make apparent that its 
nature and the properties conform, apart from minor variation of detail, 
to the description in  Sec. 2.1 of the 
standard method of construction of $L$-operators.

\subsection{The Lie algebras $d_n$}
 
We begin by treating the case
of $d_4=so(8)$. The positive roots of $d_4$ are given~\cite{hum} by
$l_i \pm l_j$, for $1 \leq i < j \leq 4$, where the $l_i$ denote the
standard basis vectors of $\R^4$ (and similarly for $d_n \;$). For the
simple roots we take
\begin{equation} \label{2.31}
\bo{r}_1=(1,-1,0,0) \; , \; \bo{r}_2=(0,1,-1,0) \; , \;
\bo{r}_3=(0,0,1,-1) \; , \; \bo{r}_4=(0,0,1,1) \quad . \end{equation}
Then for the remaining positive roots we have
\begin{equation} \label{2.32}
\bo{r}_{12}=(1,0,-1,0) \; , \;
\bo{r}_{23}=(0,1,0,-1) \; , \;
\bo{r}_{123}=(1,0,0,-1) \; , \;
\bo{r}_{24}=(0,1,0,1) \quad , \end{equation}
\begin{equation} \label{2.33}
\bo{r}_{124}=(1,0,0,1) \; , \;
\bo{r}_{234}=(0,1,1,0) \; , \;
\bo{r}_{1234}=(1,0,1,0) \; , \;
\bo{r}_{12234}=(1,1,0,0) \quad . \end{equation}
The standard form of $L$ for $d_4$ is
\begin{equation} \label{2.34} \left(
\begin{array} {cccccccc}
H_1 &&&&&&& \\
E_1 & H_2 &&&&&& \\
E_{12} & E_2 & H_3 &&&&& \\
E_{123} & E_{23} & E_3 & H_4 &&&& \\
E_{124} & E_{24} & E_4 & 0 & -H_4 &&& \\
E_{1234} & E_{234} & 0 & -E_4 & -E_3 &-H_3 && \\
E_{12234} & 0 & -E_{234} & -E_{24} & -E_{23} & -E_2 & -H_2 & \\
0 & -E_{12234} & -E_{1234} & -E_{124} & -E_{123} & -E_{12} & -E_1 & -H_1 
\end{array} \right) \quad .
\end{equation}
As before $h_r, \; r \in \{1,2,3,4 \} \; $ and the $e_{\pm \alpha}$
can be read off, their Lie algebra relations checked and seen to be
correct. So also can the $x_i$ be obtained. They satisfy (\ref{1.2})
and (\ref{2.21}), where, for $M$, we use  the $8 \times 8$ analogue of 
(\ref{2.22}). But we see
that the nice pattern of assignments of generators to places in $L$
has been modified for $d_4$, not only
because $E_4$ appears below the first main sub-diagonal, but also in
other respects.  Much the same applies to general $d_n$.

To see exactly that and how (\ref{2.34}) and its $2n \times 2n$
generalisation for $L$ for $d_n$ conform to the pattern observed above
for $c_n$ and $b_n$, we recall the Dynkin diagram of $d_n$ has a line
of linked points labelled $1,2,\ldots(n-3)$ leading to the node
$(n-2)$ from which separate the two lines of the fishtail to the
points $(n-1)$ and $n$.  For $d_4$, above, it is the fishtail roots
$\bo{r}_3$ and $\bo{r}_4$ are somehow responsible for obscuring 
the standard nature of the display (\ref{2.34}) for $L$,
and thereby the correct method of generalising it. One way to
find a suitable analogue of previous patterns locked within
(\ref{2.34}) is simply to leave out row 5 and column 5, 
which reveals a reduced display given {\sl schematically} by
\begin{equation} \label{2.35} \left(
\begin{array} {ccccccc}
x &&&&&& \\
1 & x &&&&& \\
12 & 2 & x &&&& \\
123 & 23 & 3 & x &&& \\
1234 & 234 & 0 & 4 & x && \\
12234 & 0 & 234 & 24 & 2 & x & \\
0 & 12234 & 1234 & 124 & 12 & 1 & x 
\end{array} \right) \quad .
\end{equation}
One could alternatively have left out row and column 4 from (\ref{2.34}).  
All the features seen in previous cases are apparent here.
The zero arises here because $\bo{r}_3+\bo{r}_4$ is not a root, just
as zeros in (\ref{2.20}), (\ref{2.23}) and (\ref{2.34}) reflect the
fact that if $\bo{r}$ is a root then $2\bo{r}$ is not. The schematic nature of 
(\ref{2.35}) should be emphasised. It does not, in the present case, even 
formally coincide with an $L$-operator, but it does unambiguously indicate
how the required $L$-operator is to be written down: it tells which 
$E_{\alpha}$ must go in which place, leaving only the easy insertion of signs 
to complete the construction.

For the case of $d_3  \cong a_3=su(4)$, the standard form for $L$ is
\begin{equation} \label{2.36} \left(
\begin{array} {cccccc}
H_1 &&&&& \\
E_1 & H_2 &&&& \\
E_{12} & E_2 & H_3 &&& \\
E_{13} & E_3 & 0 & -H_3 && \\
E_{123} & 0 & -E_3 & -E_2 & -H_2 & \\
 0 & -E_{123} &-E_{13} & -E_{12} & -E_1 & -H_1 
\end{array} \right) \quad , \end{equation}
which corresponds, as the above discussion requires, to the `reduced display'
\begin{equation} \label{2.37} \left(
\begin{array} {ccccc}
x &&&& \\
1 & x &&& \\
12 & 2 & x && \\
123 & 0 & 3 & x & \\
0 & 123 & 13 & 1 & x 
\end{array} \right) \quad .
\end{equation}
\subsection{Mention of $g_2$}

Finally we refer to a construction~\cite{ajmg2}  of an $L$-operator for $g_2$ 
that is based on the fact that $g_2$ is  a non-symmetric
subalgebra of $b_3$.

We here exhibit the lower triangular form of
$L$ in schematic form, 
\begin{equation} \label{2.40} \left(
\begin{array} {ccccccc}
x  &&&&&& \\
1 & x &&&&& \\
12 & 2 & x &&&& \\
112 & 12 & 1 & x &&& \\
1112 & 112 & 0 & 1 & x && \\
11122 & 0 & 112 & 12 & 2 & \phantom{x} x & \\
0 & 11122 & 1112 & 112 & 12 & 1 & \phantom{xx} x
\end{array} \right) \quad .
\end{equation}
The pattern of the places of the non-simple roots relative to the
simple ones here, despite its non-trivial nature, 
conforms very closely to that found for the $c_n$ and $b_n$
families.  We refer to~\cite{ajmg2} for explanation and full detail.

%
\section{Explicit results for the defining representations}
\label{sec_calculate}
%

\subsection{Results for $c_n$}

We begin with the important fact that, in virtue of (\ref{1.2}) and
(\ref{2.8}), the matrices $x_i$ possess the completeness relation
\begin{equation} 
\label{3.1}
  x_{i \, ab} \, x_{i \, cd}= (x_i \otimes x_i)_{ac,bd}=
  \delta_{ad} \, \delta _{cb} - J_{ac} \, J_{bd}  
 =P_{ac,bd}-K_{ac,bd} \quad . 
\end{equation}
The indices $i,a$ here vary respectively through ranges from
$1$ to $n(2n+1)=\dim \, c_n$, $2n =\dim \, \cl{V}$, which for $c_2$
is equal to $10$ and $4$. Also $P$ is the permutation operator, and $K$, 
to within a constant factor, projects onto the trivial representation in the
decomposition of $\cl{V} \otimes \cl{V}$. 

However in many contexts to which the algebras $c_n$ relate, it is
necessary to introduce the number $n(2n-1)-1=(2n+1)(n-1)$ (equal to 5
for $c_2$) of matrices $y_{\alpha}$ that complete the basis -- 
an $su(2n)$ basis in fact -- of all
traceless hermitian $2n \times 2n$ matrices. We define them so that
\begin{equation} 
\label{3.2} 
  {y_\alpha}^\dagger =y_\alpha,\quad
  \tr y_\alpha=0,\quad\tr y_\alpha y_\beta=2\delta_{\alpha \beta},\quad
  \tr x_i y_\alpha=0\quad , 
\end{equation}
and 
\begin{equation} 
\label{3.3} 
  J y_{\alpha} J^{-1}={y_{\alpha}}^T \quad . 
\end{equation}
To account for the number of the $y_\alpha$, we note that there are
$n(2n+1)$ symmetric matrices $Jx_i$ 
and $n(2n-1)$ antisymmetric ones: $Jy_\alpha$ and $J$ itself.

Since $x_i x_j+x_jx_i$ is converted to its transpose under
conjugation by $J$, it follows that we can write
\begin{equation} \label{3.4} x_i x_j+x_jx_i=\frack{2}{n}
\delta_{ij}+d_{ij\alpha} 
y_\alpha \quad . \end{equation}Given an explicit choice of the $y_{\alpha}$,
(\ref{3.4}) defines the ad-invariant tensor $d_{ij\alpha}$ so that
$d_{ii\alpha}=0$, as well as $d_{ji\alpha}=d_{ij\alpha}$. Thus we have
the product law
\begin{equation} \label{3.5} x_i x_j =\frack{1}{n} \delta_{ij} + \frack{1}{2}i
 c_{ijk} x_k + 
\frack{1}{2} d_{ij\alpha} y_{\alpha}\quad , 
\end{equation}so that 
\begin{equation} \label{3.6} ic_{ijk} = \tr x_ix_jx_k,\quad
  d_{ij\alpha} = \tr x_i x_j y_\alpha\quad . 
\end{equation}
Since the $x_i$ and $y_{\alpha}$ have the same trace properties
(\ref{1.2}) as a standard set of $2n \times 2n$ Gell-Mann matrices
$\lambda_A$ of $a_{2n-1}=su(2n)$, their well-known~\cite{MSW}
completeness relation yields, upon use of (\ref{3.1}),
\begin{equation} \label{3.7} y_{\alpha \, ab} \, y_{\alpha \, cd}  = 
\delta_{ad} \,
\delta_{cb}+J_{ac} \, J_{bd}- \frack{1}{n} \delta_{ab} \, \delta_{cd} 
= (P+K-\frack{1}{n}\,I)_{ac,bd} \quad ,
\end{equation}
where $P$ and $K$ are as in (\ref{3.1}).
It is obvious that alongside (\ref{3.5}), we should write also
\begin{eqnarray} \label{3.8} x_i y_\alpha & = & \frack{1}{2}i 
h_{i\alpha \beta} y_\beta +\frack{1}{2} d_{ij\alpha} x_j\quad , \nonumber \\ 
y_\alpha y_\beta & = & \frack{1}{n} \delta_{\alpha \beta} + \frack{1}{2}i 
h_{i\alpha \beta} x_i +\frack{1}{2} d_{\alpha \beta \gamma} y_{\gamma} \quad .
\end{eqnarray} 
These are justified by consideration of behaviours under conjugation
with $J$, the requirement that $d_{\alpha \alpha \gamma}=0$, and by
noting that the taking of traces explains why certain tensors appear
twice in the three product laws.

One of the roles of the tensor $d_{ij\alpha}$ arises in the need to
define Casimir operators of degree four and higher~\cite{dAMMPB,MP}.
For $c_2$ for example, alongside $\cl{C}_2=X_i \, X_i= \frac{1}{2}\tr
L^2$, we have 
\begin{equation} 
\label{3.9} 
  \cl{C}_4= \frack{1}{4} \mbox{Tr} L^4=t_{ijkl} X_i X_j X_k X_l\quad , 
\end{equation}
where $t_{ijkl}$ is a totally symmetric ad-invariant fourth rank
tensor.  If one wishes to deal only with quantities which carry $c_2$
vector indices, then the argument of~\cite{dAPB} is used to show that
$x_{(i} x_j x_{k)}$, where the round brackets denote symmetrisation
with unit weight over all indices enclosed by them, is
converted into minus its transpose by conjugation with $J$. Thus one
can write
\begin{equation} \label{3.10} x_{(i} x_j x_{k)} =v_{ijkl} x_k \quad , \end{equation}
thereby defining a tensor which \emph{is} totally symmetric
\begin{equation} \label{3.11} v_{ijkl}= \tr x_{(i} x_j x_{k)} x_l =
 \tr x_{(i} x_j x_k x_{l)} \quad , 
\end{equation}
and which can serve in the role of $t_{ijkl}$ in (\ref{3.9}).

But $v_{ijkl}$ can be expressed in terms of the $d_{ij\alpha}$. Thus
insert (\ref{3.4}) into
\begin{equation} \label{3.12} x_{(i} x_j x_{k)}=x_i \{ x_j \, , \, x_k \} +
x_j \{ x_k \, , \, x_i \} +x_k \{ x_i \, , \, x_j \} \quad , \end{equation}and
use (\ref{3.5}) and (\ref{3.8}). This produces a set of unwanted terms
involving the $y_\alpha$, which vanish by use of a suitable and
obvious Jacobi identity, allowing the explicit identification
\begin{equation} \label{3.13} v_{ijkl}=\frack{1}{n} \delta_{(ij} \delta_{k)l}
+\frack{1}{4} 
d_{\alpha (ij} \, d_{k)l\alpha} \quad , \end{equation}
which too is totally symmetric.
 
To go further for example for $c_3$ to build a sextic Casimir, one can
repeat the argument of~\cite{dAPB} for a totally symmetrised five-fold
product of matrices $x_i$, or use the ad-invariant totally symmetric
sixth rank tensor
\begin{equation} \label{3.14} d_{\alpha \beta \gamma} d_{(ij}{}^{\alpha} 
d_{kl}{}^{\beta}
d_{pq)}{}^{\gamma} \quad . \end{equation}
Here we raised certain indices to exempt them from the symmetrisation effect of
the round brackets, which is trivial from a metric point of view.

One use of the $h_{i\alpha \beta}$ is to define a set $\cl{R}$ of
matrices $\left( R_i \right)_{\alpha \beta}=-ih_{i\alpha
\beta}$. These define a representation $\cl{R}$ of $c_n$ of dimension
$(2n+1)(n-1)$. To prove this one uses a simple rearrangement of the
ordinary Jacobi identity of $xyy$ type, to obtain
\begin{equation} \label{3.14A} {[} R_i \, , \, R_j {]} =ic_{ijk} R_k \quad . \end{equation}
For $c_2$, $\dim \, R_i=5$, so that $\cl{R}$ is the vector representation
of $b_2 =so(5) \cong c_2$.

Turning next to identities of various sorts, we use completeness
relations to get
\begin{equation} \label{3.15} x_i \, x_i = (2n+1) \quad , \quad 
y_\alpha y_\alpha =
\frack{1}{n} \, (2n+1) (n-1) \quad , \end{equation}
and
\begin{eqnarray} \label{3.16} x_i x_j x_i= -x_j & , & y_\alpha x_i y_\alpha =
(1-\frack{1}{n})\, x_i
\quad , \nonumber \\
x_i y_\alpha x_i = y_\alpha  & , & y_\alpha y_\beta y_\alpha
=-(1+\frack{1}{n}) y_\beta 
\end{eqnarray} 
Another important class of identities includes the following
\begin{eqnarray} \label{3.17} c_{ijk} \, c_{ijl} & = & 4(n+1)
\delta_{kl} \quad ,\nonumber \\
d_{ij\alpha} \, d_{ij\beta} & = & 4(n+1) \delta_{\alpha \beta} 
\quad , \nonumber \\
d_{ij\alpha} \, d_{ik\alpha} & = & \frack{4}{n} (n^2-1) 
\delta_{jk} \quad , \nonumber \\
h_{i\alpha \beta } \, h_{j\alpha \beta} & = & 4(n-1) \delta_{ij} \quad , 
\nonumber \\\
h_{i\alpha \beta } \, h_{i\alpha \gamma} & = & 4n \delta_{\beta \gamma} \quad ,
\nonumber \\
d_{\alpha \gamma \mu} \, d_{\beta \gamma \mu} & = & \frack{4}{n} (n-2)(n+1) 
\delta_{\alpha \beta} \quad . \end{eqnarray} 

All these identities can be proved directly by the same method
(although it is easier to get the third and the fifth from their
predecessors).  For the second, one uses the definition of
$d_{ij\gamma}$ as a trace, followed by (\ref{3.1}) and elementary
properties of the $y_\alpha$.  One consequence of (\ref{3.17}) and
(\ref{3.5}) is the formula giving the $y_\alpha$ in terms of the $x_i$
\begin{equation} \label{3.17A}
2(n+1) y_\alpha = d_{ij\alpha} x_i x_j \quad . \end{equation}

A similar approach (in the proof of which (\ref{3.16}) is very useful)
works also for identities like
\begin{eqnarray} \label{3.18} 
c_{piq} \, c_{qjr} \, c_{rkp} & = & -2(n+1) c_{ijk} \quad \nonumber \\
c_{pqk} \, d_{ip\alpha} \, d_{jq\alpha} & = & \frack{2}{n} \, (n+2)(n-1)
c_{ijk} \quad . 
\end{eqnarray} 

There are many like this of evident structure all proved the same way.

To finish the discussion of identities, one recalls the set of Jacobi-type
identities listed as first class identities in~\cite{MSW} and repeated
in~\cite{MP}. There are many possibilities: $xxx$, $xxy$, $xyy$ and
$yyy$ versions for each type.  Two of these have been used, as needed,
above. Another one gives rise to the useful result
\begin{equation} 
\label{3.19} 
c_{ijl} \, c_{kml} = \frack{4}{n} \, (\delta_{ik}\delta_{jm}-
\delta_{im}\delta_{jk}) +d_{ik\alpha} \, d_{jm\alpha}-
d_{im\alpha} \, d_{jk\alpha} \quad . 
\end{equation} 
And for second class identities -- those that depend on the use of
characteristic equations -- see~\cite{MP}.

Finally, we give the eigenvalues of the Casimir operator $\cl{C}_2$ for $c_n$
for the representations $\cl{V} =(1,0,\ldots,0)$,
$\cl{R}=(0,1,0,\ldots,0)$ and $Ad=(2,0,\ldots,0)$, the adjoint representation
with matrices $(F_i)_{jk} =-ic_{ijk}$. We need $\; x_i x_i
\;$, $\; (R_i
\, R_i)_{\beta \gamma}=h_{i\alpha
\beta} h_{i\alpha \gamma} \; $ and $\; (F_i \, F_i)_{kl}= c_{ijk} \, 
c_{ijl} \; $. Using (\ref{3.15}) and (\ref{3.17}), 
it follows that $\cl{C}_2$ has 
eigenvalues $(2n+1), 4n$ and $4(n+1)$ for $\cl{V}$ , $\cl{R}$ and $Ad$.
These results agree with general formulas for the eigenvalues of Casimir 
operators given in~\cite{okubo}.
 
\subsection{Results for $so(N)$}

We may use for $\cl{D}$ matrices $x_i$ which obey (\ref{1.2}) and
$Mx_iM^{-1}=-{x_i}^T$ for all $N$, unifying the discussion and
properties of tensors for $b_n=so(2n+1)$ and $d_n=so(2n)$. So
$\dim \, x_i=N$.
Since $(Mx_i)^T=-Mx_i$, we can introduce also matrices
$y_\alpha$ such that $My_\alpha$, like $M$ itself, is symmetric. One needs a
number $\frack{1}{2} (N-1)(N+2)$ of them. In fact, it is simpler to pass 
from the 
representation of these used so far to an equivalent one with antisymmetric
matrices ${\hat{x}}_i$ and symmetric matrices ${\hat{y}}_\alpha$.
We do this here without supplying any hats. The change of
representation does not affect the values or the properties of the tensors
which enter the product laws below, but simplifies the derivation of the 
properties. These laws are
\begin{eqnarray} \label{3.51} x_i x_j & = & \frack{2}{N} \delta_{ij} + \frack{1}{2}i 
c_{ijk} x_k + \frack{1}{2} d_{ij\alpha} y_{\alpha} \quad , \nonumber \\
x_i y_\alpha & = & \frack{1}{2}i 
h_{i\alpha \beta} y_\beta +\frack{1}{2} d_{ij\alpha} x_j\quad , \nonumber \\ 
y_\alpha y_\beta & = & \frack{2}{N} \delta_{\alpha \beta} + \frack{1}{2}i 
h_{i\alpha \beta} x_i +\frack{1}{2} d_{\alpha \beta \gamma} y_{\gamma} \quad .
\end{eqnarray} 
Since the discussion follows the pattern of Sec. 3.1
closely, and the notation conforms to that
used there, we simply list results.  Completeness relations:
\begin{equation} 
\label{3.52}
x_{i \, ab} \, x_{i \, cd}= (x_i \otimes x_i)_{ac,bd}=
\delta_{ad} \, \delta _{cb} - \delta_{ac} \, \delta_{bd} 
=(P-Q)_{ac,bd} \quad , 
\end{equation}
\begin{equation} 
\label{3.53} y_{\alpha \, ab} \, y_{\alpha \, cd}  = \delta_{ad} \,
\delta_{cb}+\delta_{ac} \, \delta_{bd}- \frack{2}{N} \delta_{ab} \,
\delta_{cd} = (P+Q-\frack{2}{N}I)_{ac,bd} \quad , 
\end{equation}
where, as before,  $P$ is the permutation operator, and $Q$ is the analogue 
for $so(N)$ of the operator $K$ used above, (\ref{3.1}), for $c_n$.
Analogues of (\ref{3.15}):
\begin{equation} 
\label{3.54} 
x_i \, x_i = (N-1) \quad , \quad y_\alpha y_\alpha =
\frack{1}{N} \, (N+2) (N-1) \quad . 
\end{equation}
Analogues of  (\ref{3.16}):
\begin{eqnarray} 
\label{3.55} 
x_i x_j x_i= x_j & , & y_\alpha x_i y_\alpha =
-(1+\frack{2}{N})\, x_i
\quad , \nonumber \\
x_i y_\alpha x_i =-y_\alpha  & , & y_\alpha y_\beta y_\alpha
=(1-\frack{2}{N}) y_\beta \quad .
\end{eqnarray} 
Analogues of (\ref{3.17}):
\begin{eqnarray} 
\label{3.56} 
c_{ijk} \, c_{ijl} & = & 2 (N-2)
\delta_{kl} \quad ,\nonumber \\
d_{ij\alpha} \, d_{ij\beta} & = & 2 (N-2) \delta_{\alpha \beta} 
\quad ,\nonumber \\
d_{ij\alpha} \, d_{ik\alpha} & = & \frack{2}{N} \, (N^2-4)
\delta_{jk \quad ,} \nonumber \\
h_{i\alpha \beta } \, h_{j\alpha \beta} & = & 2(N+2) \delta_{ij} \quad ,
\nonumber \ \\
h_{i\alpha \beta } \, h_{i\alpha \gamma} & = & 2N \delta_{\beta \gamma} 
\quad , \nonumber \\
d_{\alpha \gamma \mu} \, d_{\beta \gamma \mu} & = & \frack{2}{N} \, (N-2)(N+4) 
\delta_{\alpha \beta} \quad . 
\end{eqnarray} 
The discussion of fourth rank totally symmetric tensors applies here
with obvious minor changes.  The matrices of the representation
$(2,0,\ldots ,0)$ of dimension $\frac{1}{2} (N+2)(N-1)$ can be defined
here too using $(S_i)_{\alpha \beta}=-ih_{i\alpha \beta}$. We note
here that $(2,0,\ldots ,0)$ is not the adjoint representation of
$so(N)$ but that $(0,1,0,\ldots,0)$ is. This is in contrast to the situation 
for $c_n$ for which $(2,0,\ldots ,0)$ is adjoint, and for which the 
coefficients $h_{i\alpha\beta}$ of (\ref{3.8}) define the representation
$(0,1,0,\ldots,0)$, denoted $\cl{R}$ above.

Finally one observes that results for $c_n$ translate into those for
$so(N)$ by means of the substitution $2n=N \mapsto -N$ accompanied, 
often, by a change of sign. This is of course not a new observation
-- see~\cite{cvit} -- although the sets of identities given
are. Another matter discussed in~\cite{cvit} `explains' the last
remarks of the previous paragraph. For further discussion of such matters and
many other related ones, see also the recent paper~\cite{kw}.

%
\section{Special oscillator representations}
\label{sec_oscillator}
%

\subsection{The metaplectic representation $\cl{M}_n$ of $c_n$}

Let $a_\mu$, $\mu\in\{1,2,\ldots,n\}$, denote a set of bosonic annihilation 
operators such that
\begin{equation} \label{4.1} [a_\mu \, , \, {a_\nu}^\dagger ]= \delta_{\mu\nu} \quad . \end{equation}
Then for $c_n$ we form the vector $v$ with components $v_a$,
$a\in\{1,2,\ldots,2n\} $ given by
\begin{equation} \label{4.2}
v^T= ( {a_1}^\dagger,\ldots,{a_n}^\dagger,a_n,\ldots,a_1 ) \quad , \end{equation}
and define the operators
\begin{equation} 
\label{4.3}
  X_i = \frack{1}{2}v^Tx_i(Jv)\quad, 
\end{equation}
where $J$ is the symplectic form defined in (\ref{2.8}). It is now
easy to use (\ref{4.3}) and (\ref{4.2}), as well as (\ref{2.8}), to
show (what the notation $X_i$ in (\ref{4.3}) implies), that these
$X_i$ obey the same commutation relations as do the $x_i$.  Since the
$x_i$ by construction give a representation of $c_n$, so also do the
operators $X_i$ of (\ref{4.3}). It acts in the Fock space of $n$ sets
of harmonic oscillator variables and constitutes the (reducible)
metaplectic representation $\cl{M}_n$ \cite{Itz,SW,MW,WM} of $c_n$ or of
$sp(2n,\R)$.  By use of the explicit forms used in section two for the
matrices $x_i$ of $c_2$ as well as (\ref{2.2}), we find the results
for $c_2$
\begin{eqnarray} 
\label{4.4} 
H_1=\frack{1}{2} \{ {a_1}^\dagger \, ,\, {a_1} \} \quad & , & \quad
H_2=\frack{1}{2} \{ {a_2}^\dagger \, ,\, {a_2} \} \quad , \nonumber \\
E_1={a_1}^\dagger a_2 \; & , & \; E_2=-\frac{1}{\sqrt{2}}
{{a_1}^\dagger}{}^2 \; , \; E_3=-{a_1}^\dagger {a_2}^\dagger \; , \;
E_4=-\frac{1}{\sqrt{2}} {{a_2}^\dagger}{}^2 \quad , \\
{E_{-1}}^{\dagger}= E_1 \quad & , & {E_{-\alpha}}^{\dagger}= -E_\alpha
\; , \; \alpha=2,3,4 \quad . \nonumber \end{eqnarray}  The hermiticity 
properties seen here reflect the fact that (\ref{4.4}) defines the 
generators of an infinite 
dimensional unitary
representation of the non-compact group $Sp(4, \R)$.  The operators
\begin{equation} \label{4.5} J_z=\frack{1}{2} (H_1-H_2) \; , \, 
J_+={a_1}^\dagger a_2 \; , \; J_-={J_+}^\dagger \quad , \end{equation}
and $\frack{1}{2} (H_1+H_2)$ span
the maximum compact subalgebra of $c_2$.  Also the non-compact raising
generators of $c_2$, namely $E_4 \, , \, E_3 \, , \, E_2$ are the
components of a standard set of spherical components $K_\mu$ of a
vector operator (in the precise sense of Racah) with respect to the
$su(2)$ algebra generated by the operators $\bo{J}$ in
(\ref{4.5}). The same applies to their adjoints $E_{-2} \, , \,
-E_{-3} \, , \, E_{-4}$. Furthermore, and notably, the components of
each of these vectors constitute a set of commuting operators, which
is crucial for building up the basis states of $\cl{M}_2$, and for
the solution of Gaudin models~\cite{gau}, and in algebraic Bethe
ansatz work~\cite{MW}.

As all the operators $X_i$ of (\ref{4.4}) are bilinear in creation and
annihilation operators, it follows that the Fock space $\cl{B}_2$ of
$\cl{M}_2$ breaks into two subspaces $\cl{B}_{2\pm}$ of even and odd
states in which the operators of $c_2$ act irreducibly. The ground
state of $\cl{B}_{2+}$ is the Fock vacuum $|0\rangle$ such that $a_1
|0\rangle =a_2 |0\rangle =0$, and $\cl{B}_{2+}$ involves states of
integral $j$, $\bo{J}^2=j(j+1)$.  These are constructed as
$su(2)$ multiplets with respect to $\bo{J}$ by action of the
(commutative) $K_\mu$ on the Fock vacuum. Indeed, states involving $p$
applications of $K_\mu$ components have $j=p$.  The situation with
regard to $\cl{B}_{2-}$ is similar; its ground states are a
$j=\frack{1}{2}$ doublet of states ${a_1}^\dagger |0\rangle$,
${a_2}^\dagger |0\rangle$, \ie\ the states $| \frack{1}{2} \;
\pm \frack{1}{2} \rangle$ in standard, $|j \; m \rangle$ notation, and
the action of the $K_\mu$ produces $su(2)$ multiplets with $j$-values
equal to one-half of an odd integer. Since $p$ applications of $K_\mu$
components to the states of the ground state multiplet produces a
state of $j=p+\frack{1}{2}$, it follows that $\cl{M}_2$ contains
exactly one multiplet of states of each of the allowed $j$-values.

Most of the previous discussion generalises easily and obviously from
the case of $c_2$ to $c_n$. The maximal compact sub-algebra of the
algebra of $c_n$ is of type $u(n)$. The non-compact `creation'
generators transform according to the representation of its $su(n)$
sub-algebra with highest weight twice that of the $n$-dimensional
defining representation, $\cl{V}$=$(1,0,\ldots,0)$ in highest
weight notation, of $su(n)$.  This does agree with the fact that the
non-compact `creation' generators of $\cl{M}_2$ belong to the vector
representation of $su(2)$. For $\cl{B}_{n+}$, we see that the Fock
vacuum is the ground state of the irreducible subspace of even
states. These are arranged into $su(n)$ multiplets of the type $(2p,
0,\ldots,0)$, which correspond to totally symmetric tensors. Turning
to $\cl{B}_{n-}$, it is clear that its ground states (lowest weight
states) constitute an $su(n)$ multiplet of states which transform
according to the representation $\cl{D}$ of $su(n)$. Application of
non-compact `creation' generators to these produces $su(n)$ multiplets
of states $(2p+1,0,\ldots,0)$. Thus $\cl{M}_n$ involves one
multiplet of states for each $su(n)$ representation that corresponds
to a totally symmetric tensor, and no others.

One may use the completeness relation (\ref{3.1}), and $\; v^TJv=-n \;
$, to show that the eigenvalue of $\cl{C}_2$ for $\cl{M}_n$ is
\begin{equation} \label{4.6} -\frack{n}{4} \, (2n+1) \quad. \end{equation}

We can see that the work of section 4.1 involves the splitting of
$c_n$, $\; c_n=u(n) \, \oplus \, n_+ \, \oplus \, n_- \;$, 
where $u(n)$ is the maximal compact
subalgebra of our non-compact realisation of $c_n$, and where $n_+$
contains the non-compact creation generators of $c_n$ and, notably, like $n_-$
is commutative. This is further exactly the same splitting as that of the
Lie algebra underlying~\cite{helg} the non-compact hermitian symmetric
space $Sp(2n,\R) \, /U(n)$.  The non-compactness of $\cl{M}_n$ follows
from the hermiticity relations seen in (\ref{4.4}). In general for
$c_n$, the generators of $n_-$ are \emph{minus} the adjoints of those
in $n_+$.

\subsection{Spinor representations of $d_n$}

Let $c_\mu$, $\mu\in\{1,\ldots,n\}$, and $\pi_\mu
={c_\mu}^{\dagger}$ denote a set of fermionic creation and
annihilation operators
\begin{equation} \label{4.11}
\{ c_\mu \, , \, \pi_\nu \} = \delta_{\mu \nu} \quad . \end{equation}
Now we form the vector $v$
\begin{equation} 
\label{4.12} 
  v^T = (c_1,\ldots,c_n,\pi_n,\ldots,\pi_1)\quad , 
\end{equation}
and define
\begin{equation} 
\label{4.13} 
  X_i= \frack{1}{2}v^T x_i(Mv)\quad , 
\end{equation}
where $M$ is as in (\ref{2.22}). It can be shown that, using
(\ref{4.11}) and (\ref{2.21}) that the $X_i$ of (\ref{4.13}) obey the
same commutation relations as do the matrices $x_i$. In this case of
course we use those constructed for $d_n$ in section 2.3 \, .

As before, we may translate (\ref{4.13}), \eg\ in the case of $d_3$,
into the explicit operator form
\begin{eqnarray} \label{4.14}
H_1=N_1-\frack{1}{2} \; & & , \; H_2=N_2-\frack{1}{2} \; , \; 
H_3=N_3-\frack{1}{2} \quad , \nonumber \\ 
E_1=c_1 \pi_2 \; & , &  \;  E_2=c_2 \pi_3 \quad , \quad E_{12}=c_1 \pi_3 
\quad , \nonumber \\
E_3=c_2 c_3 \; & , & \; E_{13}=c_1 c_3 \; , \;E_{123}=c_1 c_2 
\quad , \nonumber \\ 
E_{-\alpha} & = & {E_{\alpha}}^\dagger \; , \; \mbox{for all} \; \alpha \quad .
\end{eqnarray}  
Here $N_1=c_1 \, \pi_1$ is a fermion number operator, \etc Again use of 
fermionic anticommutation relations to do a check of any
Lie algebra relation here works correctly.  The representation of
$d_n$ just constructed acts in the Fock space $\cl{F}_n$ of $n$ Dirac
fermions of dimension $2^n$. The situation resembles closely that
described in section 4.1. The space $\cl{F}_n$ breaks up into two
subspaces $\cl{F}_{n \pm}$ of states of even and odd total fermion
number in which (\ref{4.13}) acts irreducibly because all the $X_i$
are bilinear in fermionic variables.  This also agrees with the fact
that $d_n$ has two inequivalent irreducible spinor representations of
dimension $2^{n-1}$. See \cite{CdAMPB}, which contains a detailed
description of the Fock space of dimension $2^{8}=256$ in the case $n=8$.

We note here too the splitting of the Lie algebra $d_n= u(n) \, \oplus  \, n_+
\, \oplus \, n_-$, where the $n_{\pm}$ are Abelian. This corresponds
to that of the compact hermitian symmetric space $SO(2n) /
U(n)$. Again the fact that $n_{\pm}$ is Abelian is important for the
construction of the basis of the $\cl{F}_{n \pm}$. In fact each of the
fundamental representations of $su(n)$ occurs once in $\cl{F}_n$, so
that \eg\ for $n=4$ the basis of $\cl{F}_{n+}$ contains two
singlets (the states of fermion number zero and four) and a $6=
(0,1,0)$, while $\cl{F}_{n-}$ contains $4=(1,0,0)$ and $\bar{4}
=(0,0,1)$.

\subsection{The spinor representation of $b_n$}

In this case we have to augment the $c_\mu$ and $\pi_\mu$ used in
section 4.2 by a single Majorana fermion $c$ in order to realise a
representation of $b_n=so(2n+1)$ like that of (\ref{4.13}). This is a
theme that has been developed considerably in previous
work~\cite{ajmmf}. For a striking application of such thinking, see
also~\cite{rsd} for Majorana parafermions.

The operator $c$ obeys
\begin{equation} \label{4.31}
c^\dagger=c \; ,\; c^2=\frack{1}{2} \; , \; \{ c \, , \, c_\mu \} =0 \; , \;
\{ c \, , \, \pi_\mu \} =0 \quad . \end{equation}
Forming
\begin{equation} \label{4.32}
v^T = (c_1,\ldots,c_n, \, c , \pi_n ,\ldots , 
\pi_1)  \quad , \end{equation}
we employ once more the definition (\ref{4.13}). This time it must be
checked by an independent calculation, one which uses (\ref{4.32})
and (\ref{4.31}), that the $X_i$ obey the same commutation relations as
do the $x_i$, in this case those that belong to the defining
representation $\cl{V}$ of $b_n$.

The explicit operators $X_i$ here consist of the first two lines of 
(\ref{4.14}) and in addition
\begin{eqnarray} \label{4.33}
E_3= c_3 \, c \; & , & \; E_{23}= c_2 \, c \; , 
\quad  E_{123}= c_1 \, c \; , \; \\
E_{1233}=c_1 c_3 \; & , &  \; E_{233}=c_2 c_3 \quad , \quad 
E_{12233}=c_1 c_3 \quad ,
\nonumber 
\end{eqnarray}  together  with hermitian conjugation relations.
Checks of their commutation relations using (\ref{4.11}) and (\ref{4.31}) do 
work.

In contrast to what was seen in section 4.1 and 4.2, in the splitting
$\; b_n =u(n) \, \oplus \, n_+ \, \oplus \, n_- \;$, the 
$n_{\pm}$ are not commutative,
making use of the elements of $n_+$ less convenient. This should be
seen alongside the fact that there is no hermitian space with
$G=SO(2N+1)$ and $H=U(N)$.

%
\section{Quadratic equations for $L$}
\label{sec_quadratic}
%

\subsection{Case of $c_n$}
\label{sec_2cn}

We apply the completeness relation (\ref{3.1}) of the matrices $x_i$ of $c_n$ 
to $L=x_i \, X_i$, where $X_i$ is given by (\ref{4.3}). This gives
\begin{eqnarray} \label{5.1}
L_{ab} & = & \frack{1}{2} v_b (Jv)_a+\frack{1}{2} (Jv)_a v_b 
\quad , \nonumber \\
       & = & (Jv)_a v_b-\frack{1}{2} \delta_{ab} \quad . 
\end{eqnarray} 
or
\begin{equation} \label{5.2}
L=(Jv) \, v^T -\frack{1}{2} \quad . \end{equation}
Hence
\begin{equation} \label{5.3}
L^2= \frack{1}{4} -(Jv) \, v^T +(Jv) \, (v^T J v) \, v^T \quad . \end{equation}
Since $\; (v^T J v)=-n \; $ follows from (\ref{3.1}) and (\ref{3.2}), we can 
eliminate
$(Jv) \, v^T$ to obtain a  quadratic relation for $L$. It is
\begin{equation} \label{5.4}
L^2 +(n+1)L +\frack{1}{4} (2n+1) =0 \quad . \end{equation}

\subsection{Cases of $b_n$ and $d_n$}
\label{sec_2bn}

The completeness relations for the matrices $x_i$ of $b_n$ and $d_n$
take the same form (\ref{3.1}), and as above, we get
\begin{equation} \label{5.11}
L=\frack{1}{2} -(Mv) \, v^T \quad . \end{equation}
Since we find $\; v^T Mv=n \; $ for $so(2n)$ and $n+\frack{1}{2}$ for 
$so(2n+1)$, and 
hence $\frack{1}{2} N$ for all $so(N)$, it follows that for all $so(N)$
\begin{equation} 
\label{5.12}
L^2 + \frack{1}{2} (N-2)L -\frack{1}{4} (N-1) =0 \quad . 
\end{equation}
The result $\tr L =0$ is compatible with (\ref{5.12}), which
implies that the eigenvalues of $L$ are $\frack{1}{2}$ and
$-\frack{1}{2} (N-1)$, because the former occurs with multiplicity $(N-1)$.

\subsection{Some other examples}

We here compare the examples of the two previous sub-sections with
some $a_{n-1}=su(n)$ examples some of which employ the definition
(\ref{1.5}).

We look first at the case of $su(2)$, setting $\gamma=\frac{1}{2}$ in
(\ref{1.5}) and using $x_i=\sigma_i$, where the $\sigma_i$
are Pauli matrices.

By considering the angular momentum addition problem for ${\bf J}=
{\bf J}_1+{\bf J}_2$ where $j_1=\frac{1}{2}$ and $j_2=j$ is arbitrary, it is 
seen that
\begin{equation} 
\label{5.21} 
  L=x_i X_i= \sigma_i ({\bf J}_2)_i \quad , 
\end{equation}
has eigenvalues $j$ and $-(j+1)$ corresponding
to the total angular momentum values $j\pm\frac{1}{2}$. It follows
that $L$ obeys the quadratic equation
\begin{equation} 
\label{5.22} 
(L+(j+1))\;(L-j)=0 \quad .
\end{equation}
Equation (\ref{5.22}) is compatible with $\tr L=0$, in view
of the multiplicities of its two eigenvalues. Equation (\ref{5.22})
could also have been obtained by the method of Secs. 5.1 and 5.2.

The present example differs from those of these sub-sections, in that
$L$ refers to an arbitrary representation of $su(2)$, as opposed to
specific ones albeit given in operator form. 
Indeed, for $su(2)$, each tensor product of the fundamental
with another representation decomposes into exactly two irreducible
components, so that in this case the quadratic relation is a property of the
Lie algebra rather than of the representations involved. Therefore we
have a quadratic relation~(\ref{5.22}) for each finite dimensional
irreducible representation of $su(2)$, the vector space on which it acts
being the quantum space $\cl{H}$ of some model.

We turn next to the case of $a_{n-1}=su(n)$. Setting $\gamma=\frac{1}{2}$
and $x_i = \lambda_i$, where these are the   
usual \cite{MSW} Gell-Mann $\lambda$-matrices of $su(n)$, we use (\ref{1.5}).
It is to be noted that the latter definition uses twice as many oscillator 
pairs as does our $c_n$ construction, (\ref{4.2}).

Then the method of Secs. 5.1 and 5.2 shows that
\begin{equation} 
\label{5.23} 
  L=x_i X_i =\lambda_i X_i \quad , 
\end{equation}
satisfies the equation
\begin{equation} 
\label{5.24}
  (L+1+\frac{N}{n}) \, (L+\frac{N}{n}-N)=0 \quad . 
\end{equation}
The operator $N=a_j{}^\dagger a_j$ here is the total bosonic number
operator, which enters the calculations via $A^\dagger A=N$ in the notation
of (\ref{1.5}).

Putting $n=2$ and replacing the operator $N$ by its eigenvalue $2j$
leads back to (\ref{5.22}). On the other hand, comparing this $su(2)$
example with the $c_1 \cong su(2)$ representation $\cl{M}_1$, we note
the former used two while the latter needed only one oscillator pair
of variables.

Since the representation operators $X_i$ in eq.~(\ref{1.5}) involve
exactly one creator and one annihilator each, the bosonic number
operator $N$ forms an invariant of this representation, \ie\
$[N,X_i]=0$. Thus the oscillator representation decomposes into
invariant subspaces, one for each eigenvalue $\lambda$ of $N$. Each
subspace corresponds to the subspace of the bosonic Fock space
generated by all states of a given number $\lambda$ of quanta. Whereas
for $n=2$, each finite dimensional irreducible representation of
$su(2)$ occurs exactly once in this decomposition, for higher $n$ the
oscillator representation contains due to the bosonic operators only
representations which are totally symmetric tensor products of the fundamental
representation. For $su(n)$, these are all representations whose
highest weight is of the form $(\lambda,0,\ldots,0)$ in terms of the
fundamental weights or which correspond to a Young diagram with just
one row of length $\lambda$.

In fact the $n=3$ example deals with the direct product $(1,0)
\otimes (\lambda,0)$ where $\lambda$ is the eigenvalue of the operator
$N$. Now
\begin{equation} 
\label{5.25}
 (1,0) \otimes (\lambda ,0)= (\lambda+1,0)\oplus(\lambda-1,1)\quad. 
\end{equation}
For the latter two, the quadratic Casimir operator of $su(3)$,
which in general has eigenvalues (see \eg~\cite{KLM}, where 
a factor $\frac{1}{9}$ occurs instead of $\frac{1}{3}$ here), 
\begin{equation} 
\label{cas}
  \cl{C}_2 (\lambda , \mu)=\frack{1}{3} (\lambda^2 +\lambda \mu +\mu^2 +
  3\lambda +3\mu) \quad , 
\end{equation}
has eigenvalues $\frac{1}{3} (\lambda +1) \, (\lambda +4)$ and 
$\frac{1}{3} (\lambda +1)^2$. Thus $L=x_iX_i$ with $X_i$ given by
(\ref{1.5}) has eigenvalues $-\frack{1}{3} \lambda$ and 
$-\frack{1}{6} \, (\lambda +3)$. Of course, (\ref{5.24}) agrees with
this when restricted to $n=3$. 

\subsection{Relevance of the quadratic equation}
\label{sec_why2}

If we consider rational solutions $T(u)$ and $R(u)$ of eq.~(\ref{1.4})
that are symmetric under the Lie algebra $\g$, we expect $T(u)$ to be
of the form
\begin{equation}
\label{gen_t}
  T(u) = \sum_{j=1}^kf_j(u)P^{(j)},
\end{equation}
where $P^{(j)}$, $j\in\{1,\ldots,k\}$, are the invariant projectors
decomposing 
\begin{equation}
  \cl{V}\otimes\cl{H}=V_1\oplus\cdots\oplus V_k
\end{equation}
into irreducible components. The $f_j(u)$ in eq.~(\ref{gen_t})
denote rational functions which are to be determined from the
condition~(\ref{1.4}). 

In Sec. 5  we have shown that the operators $L$ in
our representations given by eq.~(\ref{4.3}) and~(\ref{4.13})
satisfy quadratic relations. This reflects the fact that
$\cl{V}\otimes\cl{H}=V_1\oplus V_2$ decomposes under the action of the 
invariance algebra $\g$ into two components which are the eigenspaces
of the 
quadratic Casimir operator $C^{(2)}_{\cl{V}\otimes\cl{H}}$.
The $L$-operator has the same invariant subspaces, since
\bea 
  C^{(2)}_{\cl{V}\otimes\cl{H}} 
  & = & \sum_j {(x_j\otimes\openone + \openone\otimes X_j)}^2 \nonumber \\
  & = &  \mu_1P^{(1)} + \mu_2P^{(2)} \nonumber \\
  & = & C^{(2)}_\cl{V}\otimes\openone  + \openone\otimes 
   C^{(2)}_\cl{H}+2\,L \quad . \label{X1} \ea

In such cases, we can use, for insertion into the $RTT$ equation (\ref{1.4}),
the simple ansatz
\begin{equation}
\label{t_def}
  T(u)=u\openone + \eta L
\end{equation}
with a constant $\eta$, since this incorporates the generic linear 
combination of
all (\ie\ both) invariant projectors as demanded by~(\ref{gen_t}).

The advantage of~(\ref{t_def}) is furthermore, that this is
precisely the same form of $T(u)$ as it appears for rational $su(n)$
symmetric models where an algebraic Bethe ansatz is not only
available, but also tractable.

In fact, for $c_n=sp(2n)$ \cite{MW},
\begin{equation}
\label{r_sp}
  R(u) = u\,\openone + \eta\,P - \frac{u\eta}{\eta(n+1)+u}\,K
\end{equation}
together with $T(u)$ in~(\ref{t_def}) provide a solution
of~(\ref{1.4}). Here $P_{ac,bd}=\delta_{ad}\delta_{bc}$ flips the two
tensor factors, and $K_{ac,bd}=J_{ac}J_{bd}$ projects (up to a factor)
onto the trivial representation $\cl{V} \otimes \cl{V}$, where $\cl{V}$ 
denotes the defining representation of $c_n$. 

The analogue for $\g=so(N)$ ($N=2n+1$ for $b_n$ and $N=2n$ for $d_n$) is
\begin{equation}
\label{r_so}
  R(u) = u\,\openone + \eta\,P - \frac{u\eta}{\frack{1}{2}\eta(N-2)+u}\,Q,
\end{equation}
where $Q_{ac,bd}=\delta_{ac}\delta_{bd}$ projects (again up to a factor) onto
the trivial component of  $\cl{V} \otimes \cl{V}$, where $\cl{V}$ here
refers to the defining representation of $so(N)$. The
$R$-matrices~(\ref{r_sp}) and~(\ref{r_so}) are
given in~\cite{Resh3}. They were found in the study of
models in which the quantum space $\cl{H}$ is a particular finite-dimensional
irreducible representation. 

\subsection{Quadratic relations for finite dimensional quantum spaces}

Of course, the arguments of Sec 5.4 are also
available for finite dimensional irreducible representations $\cl{H}$
of $\g$ for which $\cl{V}\otimes\cl{H}$ decomposes into exactly two
irreducible components.

For $a_n$ we can select all representations with highest weight
$(\lambda,0,\ldots,0)$ due to eq.~(\ref{5.24}). But besides this,
we have for the conjugate representations
\begin{equation}
  (1,0,\ldots,0)\otimes(0,\ldots,0,\lambda)
    =(1,0,\ldots,0,\lambda)\oplus(0,\ldots,0,\lambda-1),
\end{equation}
and similarly for the other representations whose highest weight is an
integer multiple of a single fundamental weight.

Since tensor products of $b_n$, $c_n$ and $d_n$ representations tend
to split into smaller components than their $a_n$ counterparts, one
expects fewer quadratic relations in these cases. In fact, for $b_n$,
there is only one pair of representations 
\begin{equation}
  (1,0,\ldots,0)\otimes(0,\ldots,0,1)=(1,0,\ldots,0,1)\oplus(0,\ldots,0,1).
\end{equation}
\noindent A similar decomposition is available for $c_n$, where
\begin{equation}
  (1,0,\ldots,0)\otimes(0,\ldots,0,1)=(1,0,\ldots,0,1)\oplus(0,\ldots,0,1,0)
\end{equation}
except for $c_2$, where all
\begin{equation}
  (1,0)\otimes(0,\lambda)=(1,\lambda)\oplus(1,\lambda-1)
\end{equation}
due to the isomorphism $c_2\cong b_2$. 

Finally, for $d_n$ we find quadratic relations for the decompositions
\begin{equation}
  (1,0,\ldots,0)\otimes(0,\ldots,0,\lambda)
    = (1,0,\ldots,0,\lambda)\oplus(0,\ldots,0,1,\lambda-1)
\end{equation}
and
\begin{equation}
  (1,0,\ldots,0)\otimes(0,\ldots,0,\lambda,0)
   = (1,0,\ldots,0,\lambda,0)\oplus(0,\ldots,0,\lambda-1,0)
\end{equation}
for both spinor representations of $d_n=so(2n)$. The algebra $d_3$ is
an exception to these rules and admits many more quadratic relations
due to $a_3\cong d_3$. We remark that the
$R$-matrices~(\ref{r_sp}) and~(\ref{r_so}) are found~\cite{Resh3} in
the study of these pairs of finite-dimensional
representations where the corresponding $L$ matrix
satisfies a quadratic relation.

If one could relax the requirement that the auxiliary space $\cl{V}$ 
is chosen as the
defining representation (on which much of our discussion in the
previous sections relied), there were a few more examples of quadratic
relations. For $b_n$, they are
\begin{equation}
  (0,\ldots,0,1)\otimes(\lambda,0,\ldots,0)
  = (\lambda,0,\ldots,0,1)\oplus(\lambda-1,0,\ldots,0,1).
\end{equation}
For $c_n$, there are no more results, but for $d_n$, 
\begin{equation}
  (0,\ldots,0,1)\otimes(\lambda,0,\ldots,0)
  = (\lambda,0,\ldots,0,1)\oplus(\lambda-1,0,\ldots,0,1,0)
\end{equation}
and
\begin{equation}
  (0,\ldots,0,1,0)\otimes(\lambda,0,\ldots,0)
  = (\lambda,0,\ldots,0,1,0)\oplus(\lambda-1,0,\ldots,0,1).
\end{equation}
For $d_4$ there are in addition
\begin{equation}
  (0,0,0,1)\otimes(0,0,\lambda,0)
  = (0,0,\lambda-1,1)\oplus(1,0,\lambda-1,0)
\end{equation}
and
\begin{equation}
  (0,0,0,\lambda)\otimes(0,0,1,0)
  = (0,0,1,\lambda-1)\oplus(1,0,0,\lambda-1).
\end{equation}

\subsection{Consequences for $a_n$}

One can usefully proceed further and convert the quadratic equations
for $L$ into product laws for the operators $X_i$ involved in $L$.

Look briefly first at (\ref{5.22}), which refers to $a_1$, although no
new result emerges in this case. Using the product formula for
$\sigma_i \, \sigma_j$ in (\ref{5.22}) gives ${\bf J}^2=j(j+1)$ and
$-i\epsilon_{ijk} J_i J_j=J_k$. The latter equation, and likewise its
counterparts in other cases below, contains no new information:
\begin{equation} \label{5.31}
-i\epsilon_{ijk} J_i J_j=-\frack{1}{2} i \epsilon_{ijk} {[} J_i , J_j {]}=
\frack{1}{2} \epsilon_{ijk} \epsilon_{ijl} J_l = J_k \quad . \end{equation}

In the case of (\ref{5.24}), when the $X_i$ are the generators of the
irreducible representation $(\lambda , 0, \ldots ,0)$ of $su(n)$, 
the Kronecker delta term of the
product law~\cite{MSW} for $\lambda_i \lambda_j$ leads to
\begin{equation} \label{5.32} 
\cl{C}_2=\frack{1}{2} \, {\rm Tr}\, L^2=X_i\, X_i=
\frac{1}{2} \, (n-1) \lambda 
(1+\frac{\lambda}{n}) \quad . \end{equation}
The part containing $f_{ijk}$ is treated as in the previous
example. Thus we are left with the useful result
\begin{equation} \label{5.33} d_{ijk} X_i X_j=\frac{n+2\lambda}{2n} \, 
(n-2) X_k \quad , \end{equation}
which vanishes at $n=2$.

We note first that (\ref{5.33}) enables the calculation of the
eigenvalues for $(\lambda , 0, \ldots ,0)$ of higher order Casimir
operators. For example,
\begin{equation} \label{5.35}
\cl{C}_3= d_{ijk} X_i X_j X_k=\frack{1}{4} (n-1)(n-2)\lambda 
(1+\frac{\lambda}{n}) \, (1+\frac{2\lambda}{n}) \quad , \end{equation}
which for $n=3$ reads
\begin{equation} \label{5.36}
\cl{C}_3(\lambda ,0)= \frack{1}{18} \lambda (\lambda +3)(2\lambda +3) 
\quad . \end{equation}
This agrees to within a factor due to normalisations with the result
in~\cite{KLM}
\begin{equation} \label{5.37}
\cl{C}_3(\lambda, \mu)=\frack{1}{18} (\lambda -\mu)(\lambda +2\mu+3)
(2\lambda +\mu+3)
\quad . \end{equation}

\subsection{Consequences for $c_n$}

Here, when the operators $X_i$ are those of $\cl{M}_n$ as displayed in
(\ref{4.4}), use of (\ref{3.5}) in (\ref{5.4}) gives
\begin{eqnarray} \label{5.41}
X_i X_i & = & -\frack{1}{4} \, n(2n+1) \quad , \nonumber \\
ic_{ijk} X_i X_j &= & -2(n+1) X_k \quad , \nonumber \\
d_{ij\alpha} X_i X_j & = & 0 \quad .
\end{eqnarray}  Of these, the first agrees with (\ref{4.6}), 
the second gives nothing new, but the latter does and is useful. 
It is worth remarking also that 
the operators $X_i$ in (\ref{5.41}) are represented by infinite dimensional
matrices in the Fock space of $\cl{M}_n$.
As above, one use of 
(\ref{5.41}) is to calculate the eigenvalues for $\cl{M}_n$ of the higher
order Casimir operators of $c_n$, For example, for
\begin{equation} \label{5.43}
d_{\alpha(ij} \, d_{kl)\alpha} X_i X_j X_k X_l \quad , \end{equation}
only one term under the symmetrisation brackets contributes, doing so 
according to
\begin{equation} \label{5.44} \frack{1}{3} d_{ik\alpha} \, d_{jl\alpha} 
X_i X_j X_k X_l = \frack{1}{3} d_{ik\alpha} \, d_{jl\alpha} {[}X_i ,  X_j {]} 
X_k X_l  \quad , \end{equation}
upon use of (\ref{5.41}) again. Now use of (\ref{3.18}) and (\ref{3.17})
give an expression for the eigenvalue
\begin{equation} \label{5.45} \frack{2}{3} (n^2-1) (n+2) (2n+1) \quad . 
\end{equation}
One could pursue further the question of what is the `best' definition of 
$\cl{C}_4$, and go on to $\cl{C}_r$ for higher even $r$.
 
Another use of (\ref{5.41}) arose  in the search~\cite{MW} for solutions $T(u)$
of the Yang-Baxter algebra relations when, for the quantum
space, we have taken the Hilbert space of $\cl{M}_n$. We may use
(\ref{3.1}) and (\ref{3.7}) to write
the $c_n$ invariant solution matrix $R(u)$ (\ref{r_sp}) \cite{KS,MW} 
of the Yang-Baxter equation in the form
\begin{equation} \label{5.46}
R_{13}(u)= a(u) I \otimes I +b(u) x_i \otimes x_i +c(u) y_\alpha
\otimes y_\alpha \quad . 
\end{equation} 
Then a natural strategy for passing to the
required $T(u)$ is at hand.  This replaces the auxiliary space 3 of
$R_{13}$ in (\ref{5.46}) by the quantum space $\cl{H}$, and passes from
(\ref{5.46}) to $T(u)$, via $x_i \mapsto X_i$ and $y_\alpha
\propto d_{ij\alpha}x_i\, x_j
\mapsto Y_\alpha$, with  $Y_\alpha \propto d_{ij\alpha} X_i X_j$. 
Since in the present case, the latter is seen from (\ref{5.41}) to be a 
vanishing operator, solution of (\ref{1.4}) turns
out, as might be anticipated, to be particularly simple \cite{MW}.


\acknowledgements

This work was partly supported by PPARC. F.W.~is grateful to Trinity
College, Cambridge for an IGS, and to EPSRC for a research
grant. H.P.~would like to thank DAAD for a scholarship
``Doktorandenstipendium im Rahmen des gemeinsamen
Hochschulsonderprogramms~III von Bund und L\"andern''. 


\end{document}